\documentclass{mem}
\usepackage{natbib}\usepackage{txfonts}\usepackage{balance}
\usepackage{graphicx, subfigure}
\usepackage{textcomp}
\usepackage[a4paper,breaklinks,dvipdfm]{hyperref}
\idline{75}{282}
\begin{document}
\def\kms{$\mathrm {km s}^{-1}$}

\title{
Bayesian analysis to identify very low-mass members of nearby young stellar kinematic groups
}

   \subtitle{}

\author{
J. \,Gagn\'e\inst{1},
D. \, Lafreni\`ere\inst{1},
R. \, Doyon\inst{1},
L. \, Malo\inst{1},
J. \, Faherty\inst{2},
\'E. \, Artigau\inst{1}
          }

  \offprints{J. Gagn\'e}

\institute{
D\'epartement de Physique and Observatoire du Mont-M\'egantic, Universit\'e de Montr\'eal,
C.P. 6128 Succ. Centre-ville, Montr\'eal, Qc
H3C 3J7, Canada
\email{jonathan.gagne@astro.umontreal.ca}
\and
Department of Astronomy, 
Universidad de Chile,
Cerro Calan, Las Condes
\email{jfaherty17@gmail.com}
}

\authorrunning{Gagn\'e}

\titlerunning{VLM objects in NYAs}

\abstract{
We describe our all-sky survey for $>$M4 candidate members to nearby, young associations from the 2MASS and WISE catalogs using bayesian inference. We report the first results, including 38 highly probable candidates showing spectroscopic signs of low-gravity (and thus youth). The latest of these objects would correspond to a 11~\textendash~13~M$_{Jup}$ object, around the limit of the planetary regime.
}
\maketitle{}

\section{Introduction}

Recently, many efforts have been made for finding Very Low-Mass (VLM) objects down to the planetary mass regime in young associations. Several reasons can account for this. First, the atmospheres of free-floating planetary mass objects can be easily studied as giant exoplanet analogs, because these objects would not be masked by a bright, primary star. The first such objects that have been found seem to show great diversity of spectral features, even at a fixed age and temperature, which strengthens the possibility those objects are analogs to giant, gaseous exoplanets, as well as demands for more such discoveries. Since young objects are warmer and brighter, they are also easier to study. They even provide good targets for the direct imaging of exoplanets, since such young exoplanets would also be brighter than their old counterparts, and the contrast ratio required to achieve such discoveries would be lower for a fainter primary star. Another great outcome of finding those objects would be the possibility to study the bottom of the Initial Mass Function (IMF) in a coeval environment, not to mention the possibility to improve atmospheric models of young brown dwarfs, which are still imprecise because of the lack of observations, as well as the difficulty of dealing with the large amount of dust contained in their photospheres. We have chosen to focus our search on young very low-mass stars and brown dwarfs in Nearby, Young Associations (NYAs) such as TW Hydrae (TWA; 8 - 12 Myr; \citealp{2004ARA&A..42..685Z}), $\beta$ Pictoris (BPMG; 12 - 22 Myr; \citealp{2001ApJ...562L..87Z}), Tucana-Horologium (THA; 10 - 40 Myr; \citealp{2000AJ....120.1410T}, \citealp{2001ASPC..244..122Z}), Carina (CAR; 10 - 40 Myr; \citealp{2008hsf2.book..757T}), Columba (COL; 10 - 40 Myr; \citealp{2011ApJ...732...61Z}),  Argus (ARG; 10 - 40 Myr; \citealp{2011ApJ...732...61Z}) and AB Doradus (50 - 120 Myr ; \citealp{2004ApJ...613L..65Z}). Even though these NYAs are all closer than 100~pc, most of their low-mass members ($>$K5) remain to be identified because their currently known members were uncovered in the Hipparcos mission, which is limited to relatively bright objects. Since those moving groups are close and have ages between 8~\textendash~120 Myr, they present several advantages for the task of identifying young low-mass objects: 1) their members will be even brighter and thus easier to study, 2) they are old enough so that they are no longer embedded in dust, 3) they have not significantly dispersed yet, so their members share similar Galactic position ($XYZ$) and space velocities ($UVW$) and 4) they span a significant age range, which means each NYA will serve as a benchmark for their evolution in time. Furthermore, the 10~\textendash~30 Myr range is well covered, corresponding to the period where gaseous and terrestrial planets form \citep{2003ApJ...599..342S}. However, we must overcome a significant difficulty in order to identify such new members to those NYAs: since they are close-by and have begun dispersing, they cover great portions of the sky as viewed from earth. The fact that we do not have access to parallaxes and radial velocities for most low-mass objects renders this task even more difficult. Even worse, we expect that most of the potential members won't even have been spectroscopically confirmed as brown dwarfs.\\

\begin{figure}[]
\resizebox{\hsize}{!}{\includegraphics[clip=true]{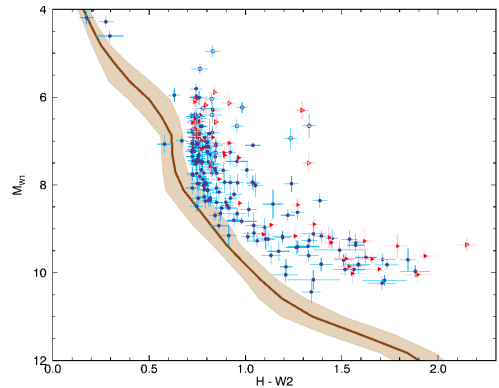}}
\caption{
\footnotesize
Color-magnitude sequence for field objects (brown line and shaded region), and our candidates with (red dots) and without (blue dots) spectroscopic confirmation of low-gravity. We can see that they are significantly redder than the old sequence, consistent with those being low-gravity objects with more dust in their photosphere.
}
\label{fig:CMD}
\end{figure}


\begin{figure*}[]
	\begin{center}
		\subfigure{
		 \includegraphics[width=0.485\textwidth]{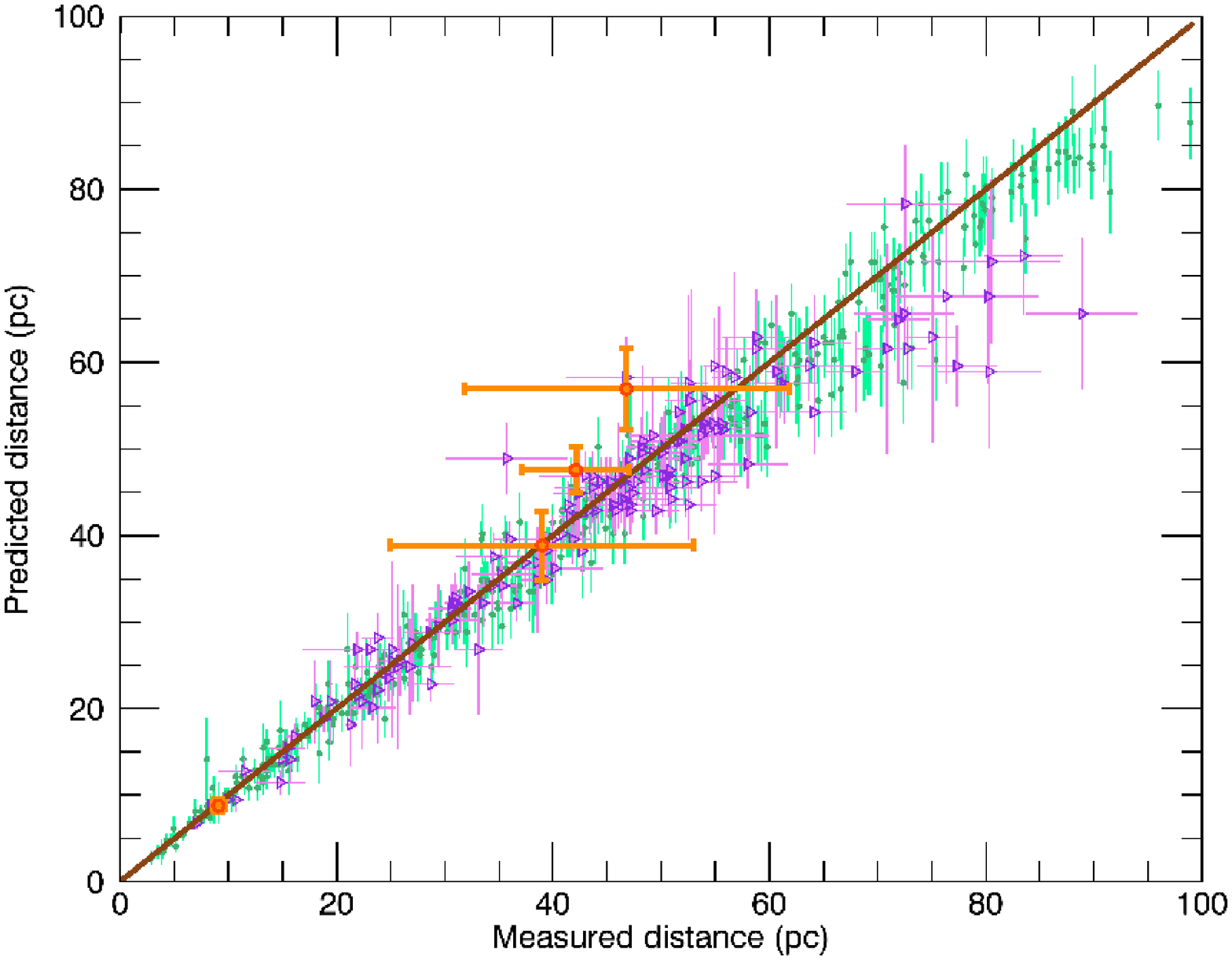}
		}
		\subfigure{
		 \includegraphics[width=0.485\textwidth]{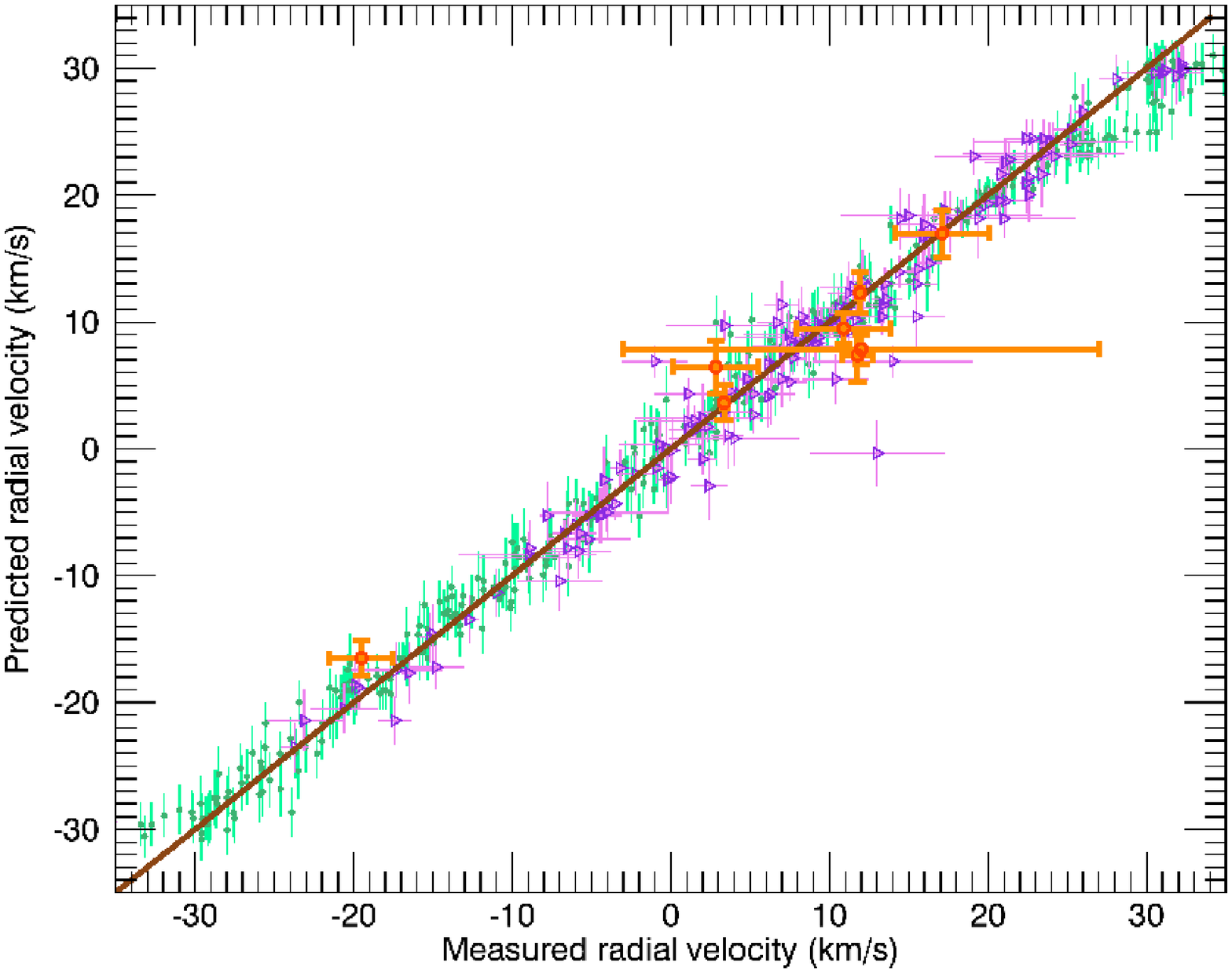}
		}
	\end{center}
\caption{
\footnotesize
Predicted distances and radial velocities for \emph{Bona Fide} members of NYAs (triangles), synthetic Montecarlo objects drawn from our NYAs models (small dots) and our candidates for which those measurements are available (thick open circles). The dark brown line represents a dependence. Our estimated errors agree well with the observed scatter. True distances (radial velocities) are retrieved within 8.0~\% (1.6~\kms).
}
\label{fig:dstat}
\end{figure*}


\section{Method}

In order to overcome the precise difficulties previously described, \cite{2013ApJ...762...88M} proposes using bayesian inference in order to identify highly probable candidates to NYAs within possibly large samples of objects for those we have at least a position and proper motion. Here, we use a slightly modified version of this method and apply it to an all-sky sample of 650 000 red objects which we have built from a correlation of the 2MASS \citep{2003yCat.2246....0C} and WISE \citep{2012yCat.2311....0C} surveys. We have limited our search outside of the galactic plane ($\left | b \right | >$ 15\textdegree) in order to avoid crowded fields. Various filters have been applied in order to reject possible contaminants such as red galaxies and giant stars, and to keep only objects whose photometry is consistent with $>$M4 objects. Applying bayesian inference to this sample has yielded more than 300 highly probable candidates to NYAs, which are currently being followed so that we can confirm them as new \emph{Bona Fide} members. Included in the bayesian inference, we have used $J$ - $K_s$, as well as $H$ - $W2$ colors as a function of absolute $W1$ magnitude in order to give a bigger weight to young hypotheses for objects which have very red colors (see Figure~\ref{fig:CMD}). \\

\begin{figure}[t!]
\resizebox{\hsize}{!}{\includegraphics[clip=true]{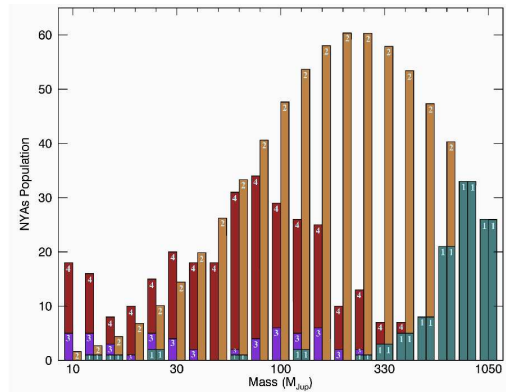}}
\caption{\footnotesize
Mass population for (1) \emph{Bona Fide} members of NYAs, (2) a fiducial log-normal IMF (described in text), candidates from this project (3) with and (4) without spectroscopic confirmation of low-gravity. We seem to find too many planetary mass candidates compared to the predictions of the IMF.
}
\label{fig:IMF}
\end{figure}

We have estimated the mass distribution of our candidates by using the predicted distance (see Figure~\ref{fig:dstat} for an accuracy assessment of these predictions) to compute 2MASS and WISE absolute magnitudes, as well as the known NYA ages in combination with the AMES-Cond isochrones \cite{2003A&A...402..701B} and the BT-SETTL atmosphere models (\citealp{2013arXiv1302.6559A}, \citealp{Rajpurohit:ta}). The resulting distribution is displayed in Figure~\ref{fig:IMF} and compared with the expected remaining members to be discovered when supposing 1) that the NYAs are approximately complete at masses around 1~M$_\odot$ and 2) that the IMF of NYAs can be approximately described as a fiducial log-normal function with $m_c$ = 0.25~M$_\odot$ and $\sigma$~=~0.5~dex (\citealp{2012EAS....57...45J}, \citealp{2005ASSL..327...41C}). We seem to be finding more planetary mass candidates than predicted with this IMF, which could be a sign that we are finding planetary-mass objects that were not formed as brown dwarfs, but rather as ejected planets. However, we must confirm these candidate members before we can draw any conclusions of this kind. We have estimated our contamination rate to be between 20~\% and 30~\% for M5~\textendash~L0.

\section{Candidates follow-up}

\begin{figure*}[]
	\begin{center}
		\subfigure{
		 \includegraphics[width=0.45\textwidth]{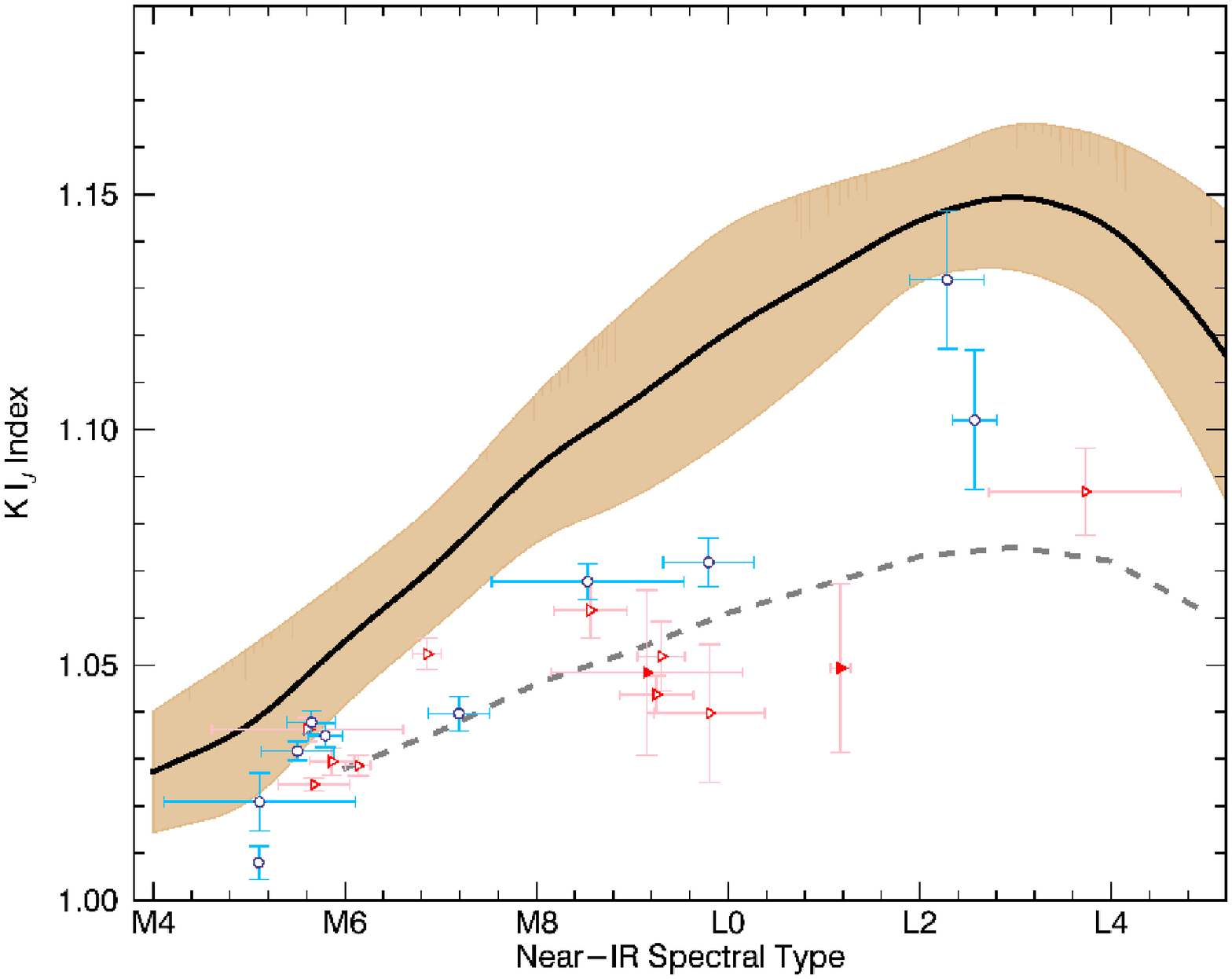}
		}
		\subfigure{
		 \includegraphics[width=0.45\textwidth]{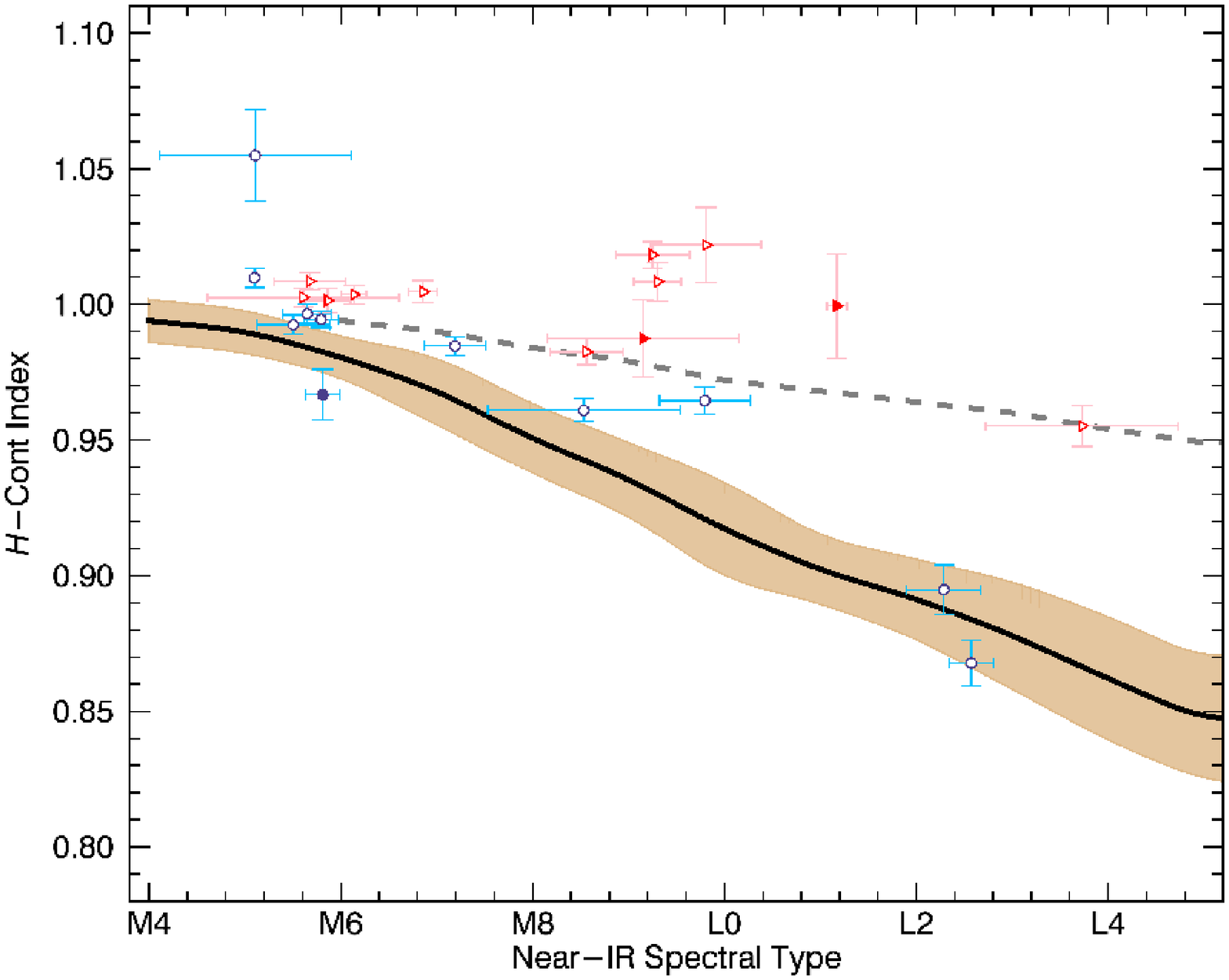}
		}
		\subfigure{
		 \includegraphics[width=0.45\textwidth]{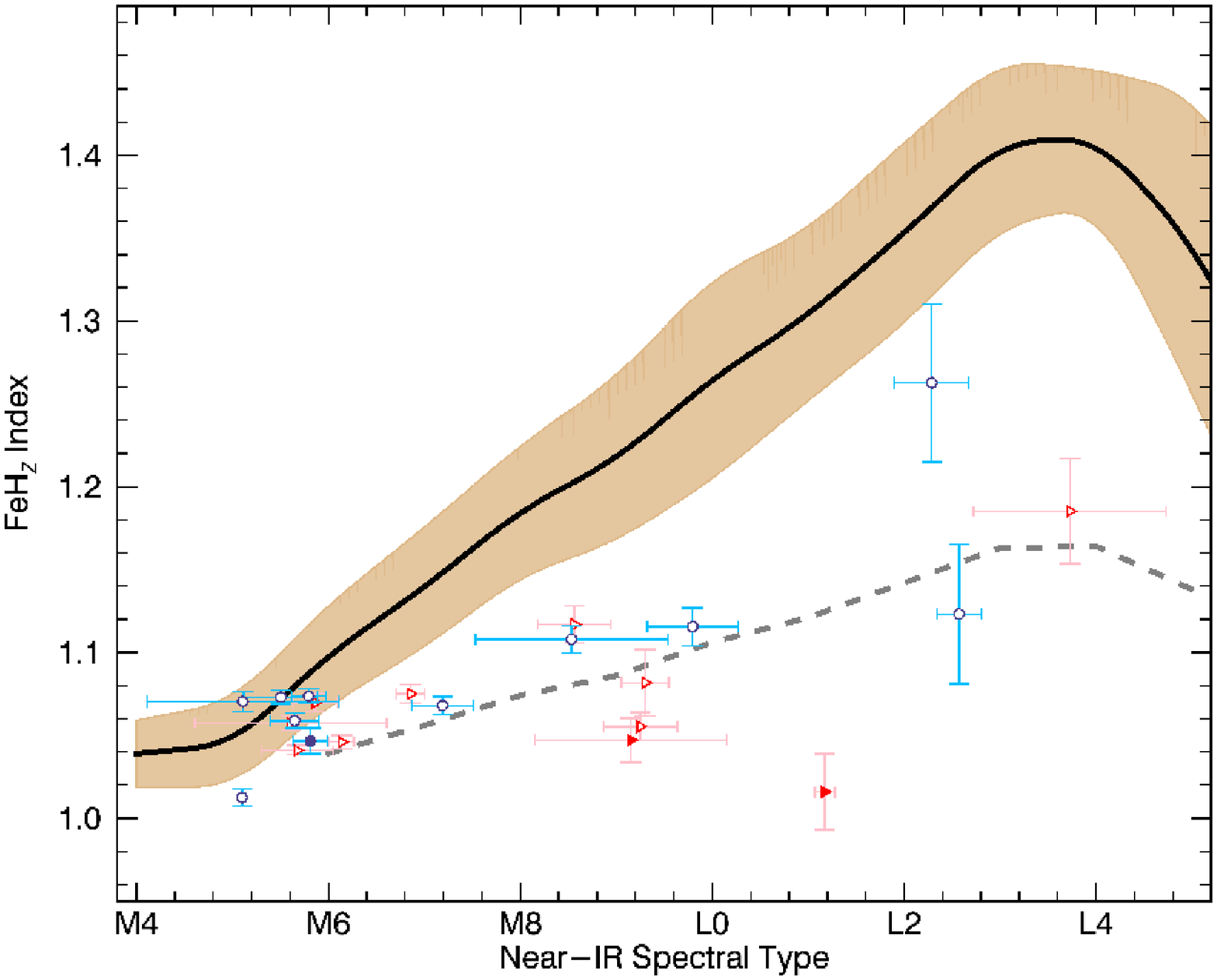}
		}
		\subfigure{
		 \includegraphics[width=0.45\textwidth]{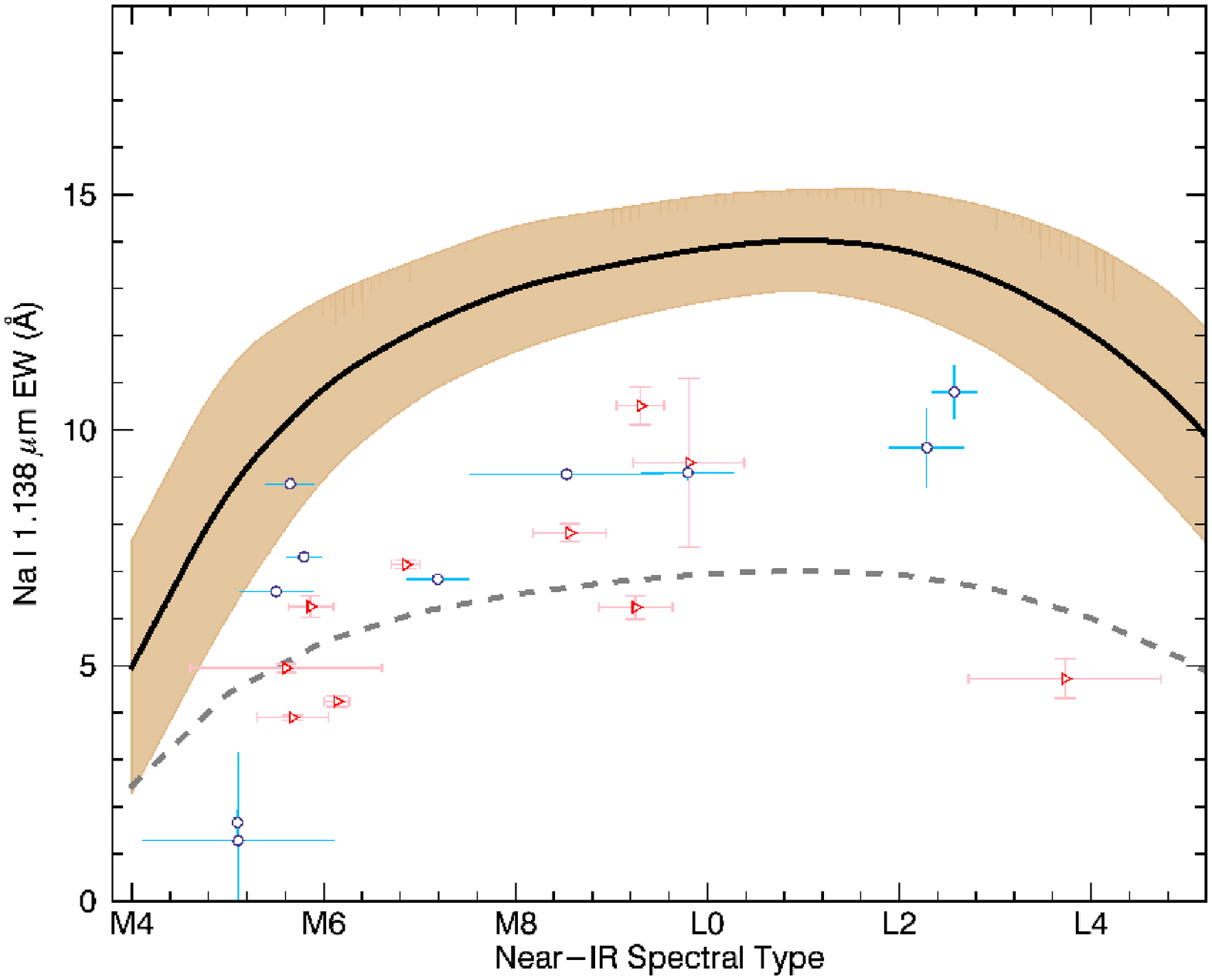}
		}
	\end{center}
\caption{
\footnotesize
K~I$_J$, FeH and $H$-cont indices defined by \cite{Allers:2013tq} and Na~I equivalent widths for our intermediate (circles) and very low gravity (triangles) candidates. The black line and beige region delimit the high gravity dwarfs sequence along with its scatter. The dotted grey line delimits the very low gravity region from the one where intermediate gravity objects typically fall. Filled symbols indicate objects with low resolution spectra (R~$\approx$~300) and open symbols indicate those with moderate resolution spectra (R~$\approx$~1200).
}
\label{fig:Allers}
\end{figure*}

\begin{figure}[]
\resizebox{\hsize}{!}{\includegraphics[clip=true]{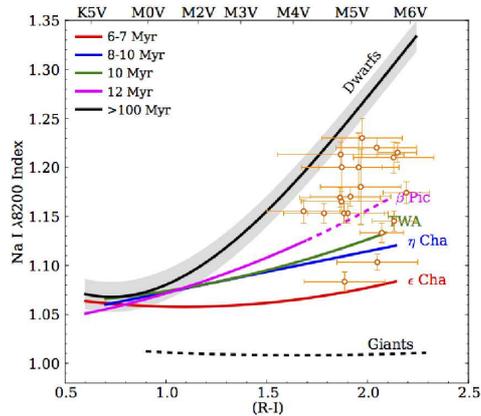}}
\caption{
\footnotesize
NaI index sequence for old dwarfs (thick line and gray region), various NYAs and low-gravity candidates from our survey (open circles), as defined by \cite{2004MNRAS.355..363L}. We can see that their relative gravity corresponds to the 6~\textendash~100 Myr age range, in agreement with the ages of NYAs considered here. Sequences are drawn from \cite{2011AJ....142..104R}.
}
\label{fig:NAI}
\end{figure}

We have used SpeX on the 3.0-m IRTF and OSIRIS on the SOAR 4.1-m telescope in order to obtain medium-resolution (R~$\approx$~1200) Near-InfraRed (NIR) spectroscopy of our brightest candidates, in order to confirm their low surface gravity (and thus youth) by using several gravity sensitive indices described in \cite{Allers:2013tq}. As a result, we have yet found 16 $>$M5 objects classified as intermediate (5) or very low (11) gravity displayed in Figure~\ref{fig:Allers}. For two of our faint candidates, we have used FIRES on the MAGELLAN 6.5-m telescope to obtain low-resolution (R~$\approx$~300) NIR spectroscopy, from which we can assess low-gravity in the same way albeit with fewer indices. Both objects turned out as very low-gravity, highly probable candidates to the TWA, which makes them potential 11~\textendash~13 M$_{Jup}$ and 13~\textendash~15 M$_{Jup}$ objects at the boundary of planetary masses. Earlier targets were followed with GMOS-S and GMOS-N on both Gemini telescopes in order to measure the equivalent width of the Na I doublet at $8174~\AA$, which is also gravity sensitive (\citealp{2004MNRAS.355..363L}, \citealp{2011AJ....142..104R}). This has yet yielded 20 M4~\textendash~M8 very strong candidates that show weaker Na I equivalent width, consistent with them being low-gravity and thus young dwarfs (see Figure~\ref{fig:NAI}). Several of our strongest bright candidates are currently being observed with CRIRES at the VLT, in order to obtain high-resolution $K$-band spectroscopy which will allow us to measure their radial velocities at a precision of 1~\kms, which is sufficient distinguish between field objects and NYAs \emph{Bona Fide} members. More observations for NIR spectroscopy are also already planned for completing the spectroscopic characterization of our whole sample, and more candidates in the $>$L0 regime are also being currently uncovered.

\section{Conclusions}

We have summarized our survey for $>$M4 members to NYAs. We show that bayesian analysis can be used to significantly cut down the initial number of candidates in an all-sky search, and then show first results from this study. It is expected that this project will unveil a significant portion of young, planetary-mass objects in NYAs, which will be easy to study in details due to their short distance and larger intrinsic brightness than old objects.

\begin{acknowledgements}
We would like to address special thanks to Adric Riedel for generously sharing valuable parallax data with our team. This work was supported in part through grants from the Natural Sciences and Engineering Research Council of Canada. This research has made use of the SIMBAD database and VizieR catalogue, operated at Centre de Donn\'ees astronomiques de Strasbourg (CDS), Strasbourg, France. 
\end{acknowledgements}
\bibliographystyle{aa}
\bibliography{Gagne}

\end{document}